\documentclass[useAMS,usenatbib]{mn2e}
\usepackage{natbib}

\usepackage{deluxetable}
\usepackage[dvips]{graphics}
\usepackage{epsfig}
\usepackage{amsmath}
\usepackage{aas_macros}

\def\mearth{{\rm\,M_\oplus}}
\def\rearth{{\rm\,R_\oplus}}

\def\gcm{\rm g \ cm^{-3}}

\title[Migration-driven diversity of super-Earth compositions]{Migration-driven diversity of super-Earth compositions}
\author[Raymond et al]{Sean N. Raymond$^1$\thanks{E-mail: rayray.sean@gmail.com}, Thibault Boulet$^\mathrm{2}$, Andre Izidoro$^3$, \newauthor Leandro Esteves$^3$, \& Bertram Bitsch$^4$\\
$^\mathrm{1}$Laboratoire d'Astrophysique de Bordeaux, CNRS and Universit{\'e} de Bordeaux, All{\'e}e Geoffroy St. Hilaire, 33165 Pessac, France \\
$^\mathrm{2}$Institut d'Astrophysique et de G{\'e}ophysique, Laboratoire d'Imagerie de syst{\`e}mes Stellaires et Plan{\'e}taires, Li{\`e}ge, Belgium\\
$^3$UNESP, Univ. Estadual Paulista - Grupo de Din{\`a}mica Orbital
Planetologia, Guaratinguet{\`a}, CEP 12.516-410, S{\~a}o Paulo, Brazil\\
$^4$Max-Planck-Institut f\"ur Astronomie, K\"onigstuhl 17, 69117 Heidelberg, Germany
}

\begin{document}

\date{Accepted to MNRAS Letters May 25, 2018}

\pagerange{\pageref{firstpage}--\pageref{lastpage}} \pubyear{2017}

\maketitle

\label{firstpage}

\begin{abstract}
A leading model for the origin of super-Earths proposes that planetary embryos migrate inward and pile up on close-in orbits. As large embryos are thought to preferentially form beyond the snow line, this naively predicts that most super-Earths should be very water-rich. Here we show that the shortest-period planets formed in the migration model are often purely rocky. The inward migration of icy embryos through the terrestrial zone accelerates the growth of rocky planets via resonant shepherding. We illustrate this process with a simulation that provided a match to the Kepler-36 system of two planets on close orbits with very different densities. In the simulation, two super-Earths formed in a Kepler-36-like configuration; the inner planet was pure rock while the outer one was ice-rich. We conclude from a suite of simulations that the feeding zones of close-in super-Earths are likely to be broad and disconnected from their final orbital radii. 
\end{abstract}


\section{Introduction}
While the demographics of close-in low-mass planets (broadly referred to as `super-Earths') have been known for some time~\citep{lissauer11b,batalha13,fabrycky14}, a handful of recent studies have focused on the closest-in planets. The atmospheres of low-mass, extremely hot planets should be entirely photo-evaporated away~\citep[e.g.,][]{lammer03,baraffe06,hubbard07b,raymond08a}. Within the `photo-evaporation valley' a planet's size must correspond to the solid (atmosphere-free) planet itself~\citep{lopez13,owen13,sanchisojeda14,carrera18}. From their inferred sizes and masses, the closest-in planets appear to be rocky, not icy~\citep{owen17,lopez17,vaneylen17,jin18}.  

The formation mode of super-Earths can in principle be inferred from a combination of the planetary system architecture and the planets' bulk compositions, as long as a rough rocky vs. icy discrimination can be made~\citep{raymond08a}. To date, seven models have been proposed to explain these planets' origins~\citep[see][]{raymond08a,raymond14}.  Two viable candidates remain: the so-called {\em drift} and {\em migration} models.

The drift model~\citep{boley13,chatterjee14,chatterjee15} proposes that a large flux of pebbles drift inward and pile up at pressure bumps, greatly enhancing the abundance of solids in the inner disk.  Most planetary growth takes place close-in and the planets are expected to be rocky.  This contrasts with the {\it in situ} accretion model, which invokes growth close-in without any enhancement of solids relative to gas or orbital migration. The in-situ accretion model is not self-consistent because planet growth is so fast that migration is unavoidable~\citep{bolmont14,inamdar15,ogihara15}. 

The migration model~\citep{terquem07,ogihara09,mcneil10,ida10,cossou14,izidoro17} proposes that Earth-mass or larger planetary embryos grow across the disk, migrate inward and pile up close to the star. The migration phase produces a chain of resonant orbits in which the innermost planet is anchored at the inner edge of the disk, where a strong positive torque balances the other planets' negative disk torques~\citep{masset06,romanova06}. When the gaseous disk dissipates, most resonant chains become unstable and the resulting systems have been shown to match the Kepler period ratio and multiplicity distributions~\citep{izidoro17}. It is generally thought that the planets formed in the migration model should be water-rich. Planetary embryos are expected to grow faster and larger beyond the snow line~\citep{kokubo02,morby15}, and it stands to reason that ice-rich embryos should be the first to migrate.  


There is a complication to the simple idea that the migration model produces ice-rich super-Earths. Icy embryos migrate through a zone of presumably rocky bodies~\citep{izidoro14}. While these rocky bodies are much less massive, the migrating icy embryo may pile up rocky material in inner resonances and in principle accelerate the growth process of rocky planets~\citep{fogg05,raymond06c,mandell07,izidoro14}. Thus, even if the migrating bodies are icy, the planets closest to the star may be purely rocky, and some rocky embryos may grow fast enough to migrate. In addition, any embryos -- including those that originate past the snow line -- are subject to water loss from $^{26}$Al heating if they form quickly~\citep{grimm93,monteux18} as well as later collisional losses~\citep{genda05,marcus10}.

In this {\em Letter} we use dynamical simulations to show that the migration model produces a broad diversity in the compositions of super-Earths. In many systems the closest-in planets are rocky. In some cases Earth-sized rocky planets survive on orbits exterior to those of ice-rich planets. The feeding zones of close-in planets are generally very wide and disconnected from their final orbital locations.  We illustrate these ideas using a simulation that produced a remarkably good analog to the Kepler-36 system of two planets on very close orbits with very different densities~\citep{carter12}. 

\section{Simulation setup}
Our simulations are based on the model of \cite{izidoro17}. We now summarize the main ingredients.

We used the disk structure and evolution model of \cite{bitsch15}. The gaseous disk evolved viscously over its 5.2 Myr lifetime, with a rapid final dissipation over the last 0.1 Myr~\citep[for details see][]{izidoro17}. Migration was calculated using the torque formulae from \cite{paardekooper10,paardekooper11} for our disk model including a correction to the corotation torque associated with the embryos' eccentricities~\citep{bitsch10,cossou13,fendyke14}. Migration may be directed either inward or outward depending on a planet's mass and its location within the disk~\citep[see][]{bitsch13,bitsch14,bitsch15}. Type 1 damping of eccentricities and inclinations~\citep{papaloizou00,tanaka04} was included using the equations of \cite{cresswell08}. The integrator is based on the {\tt Mercury} hybrid symplectic integrator~\citep{chambers99}. Collisions were treated as inelastic mergers. 

Our simulations started from a range of distributions of planetary embryos, assumed to have formed by planetesimal or pebble accretion~\citep{kokubo02,ormel10,lambrechts12,morby15}.  We included $25 - 40 \mearth$ in planetary embryos, with a division between rocky and icy bodies at 5 AU~\citep[the location of the snowline at the start of the disk's evolution][]{bitsch15}. Rocky embryos contained 20-40\% of the total embryo mass. In some cases rocky embryos were given randomly-chosen masses from $0.1-0.5\mearth$ while icy embryos had masses of $0.5-4.5 \mearth$ and in other simulations all embryos were given masses of $0.5-4.5 \mearth$. This therefore encompasses both a scenario in which no rocky embryos migration and one in which they do. Rocky embryos were placed beyond 0.7 AU; in some cases in a relatively narrow annulus and sometimes in a broad disk. The total number of embryos was 20-40, with a similar number of rocky and icy bodies. All bodies were given small random initial inclinations of up to $1^\circ$ and eccentricities of up to 0.01. For calculating radii (to determine collisions), rocky and icy embryos were given different bulk densities, in some cases of 3 $\gcm$ and 1.6 $\gcm$, respectively, and in other cases 5.5 $\gcm$ and 2 $\gcm$. In addition to the variations in initial embryo location, we changed the disk's metallicity $Z$ between 0.25\% and 1\%, which affects the disk's temperature structure and the associated migration map~\citep{bitsch15}. We ran 95 simulations, each integrated for 5-50 Myr beyond the disk's dispersal.  

\section{Formation of a Kepler-36 analog}
\begin{figure}
  \leavevmode \epsfxsize=8cm\epsfbox{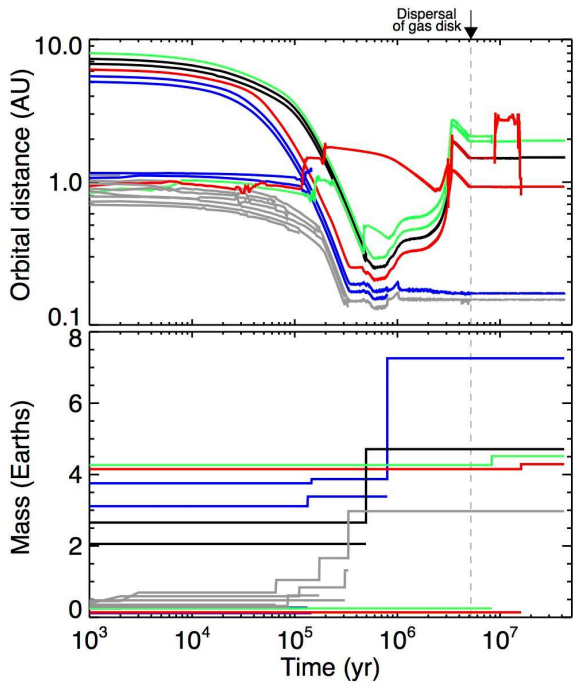}
    \caption[]{Orbital (top panel) and mass (bottom) evolution of a simulation that produced two close-in planets reminiscent of the Kepler-36 system.  All bodies incorporated into a given planet have the same color such that each planet's feeding zone can be inferred from the original orbital distance of its components.  The disk dispersed after 5.2 Myr as indicated by the dashed line. An animation of the simulation is available here: https://tinyurl.com/yaek765p.}
     \label{fig:kep36_evol}
    \end{figure}

Figure~\ref{fig:kep36_evol} shows the evolution of a particularly interesting simulation. The simulation started with six ice-rich embryos between 5 and 8.2 AU with masses from 2 to $4.2 
\mearth$ for a total of $20 \mearth$, and 13 rocky embryos from 0.7 to 1.2 AU with masses from 0.1 to $0.45 \mearth$ for a total of  $3.75 \mearth$. As ice-rich embryos migrated inward, some rocky embryos were accreted and others were scattered outward. The remainder of the rocky material was shepherded inward by inner resonances with the innermost migrating icy embryo; first by the 8:5 and later by the 3:2 and 5:4 resonances. This inward-shepherded material was compressed into a narrow radial zone and grew rapidly by collisions into a single, $3 \mearth$ rocky planet. 

\begin{figure}
  \begin{center} 
  \leavevmode \epsfxsize=8cm\epsfbox{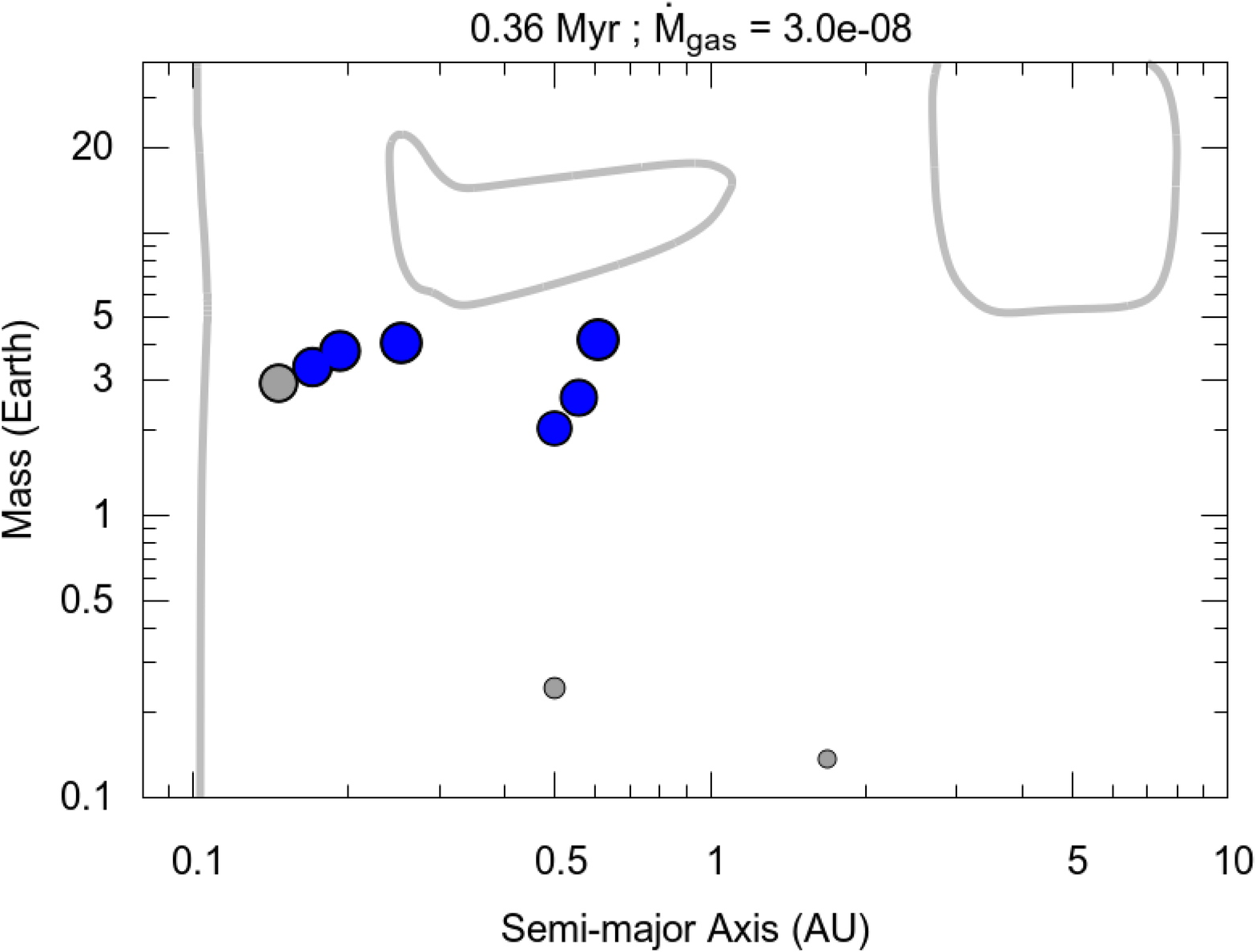}
  \leavevmode \epsfxsize=8cm\epsfbox{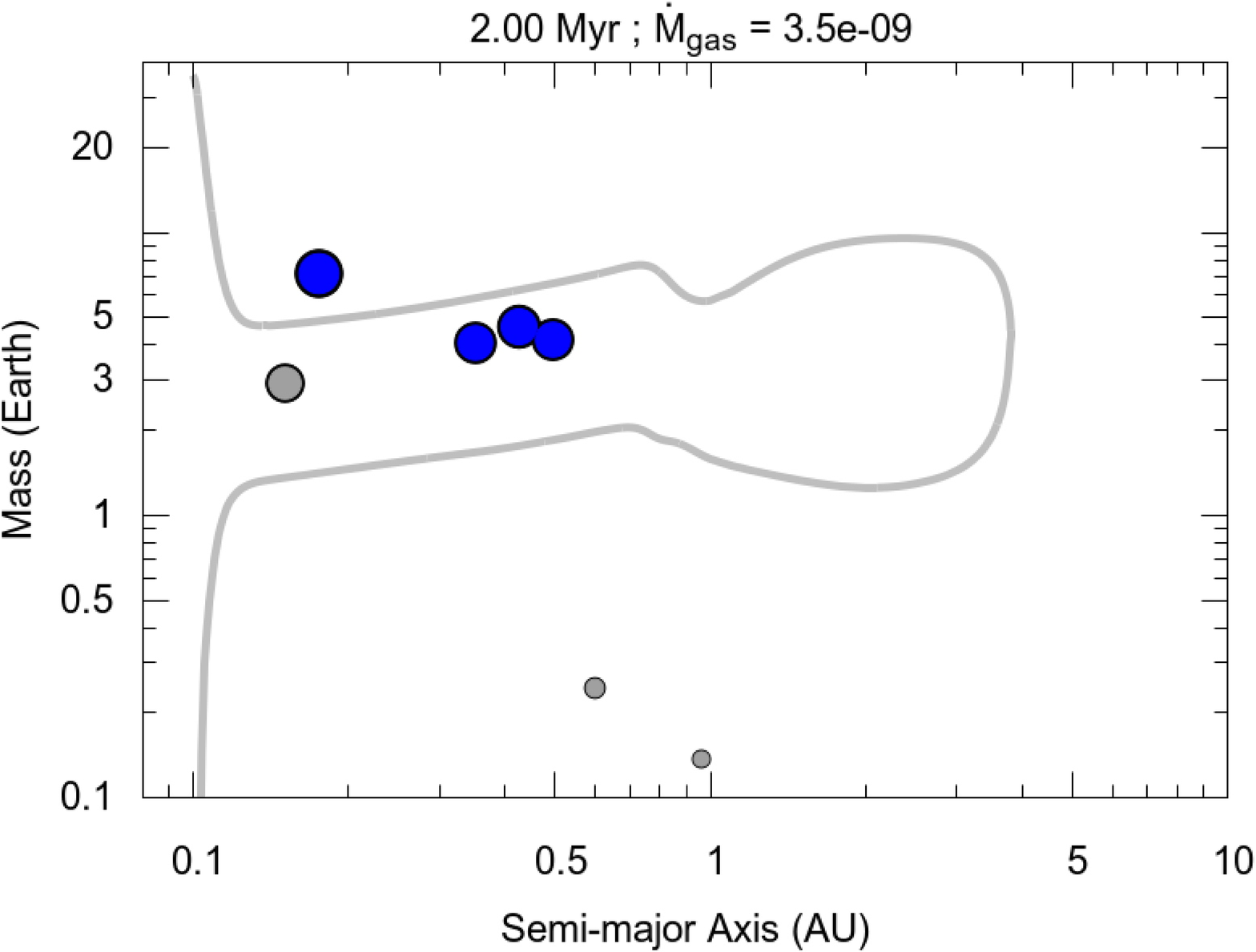}
    \caption[]{Two snapshots of the system's evolution, superimposed on the disk's migration map. The gray curves show the zero-torque (no migration) zone.  Within the gray contours (i.e., for $2-4\mearth$ planets out to $\sim3$~AU in the bottom panel) migration is directed outward. Migration is directed inward everywhere else in mass-orbital distance space.  In the top panel, all of the embryos migrate inward.  In the bottom panel, only the large inner ice-rich embryo feels a negative (inward) torque from the disk, but it holds back the outward migration of the innermost rocky embryo.  } 
     \label{fig:kep36_migmap}
\end{center}
    \end{figure}

After roughly 1 Myr, the region of outward migration in the disk shifted to encompass the growing planets due to the evolution of the disk itself (see Fig.~\ref{fig:kep36_migmap}; in this case the disk's metallicity was 0.5\%).\footnote{We note that the outward migration of embryos had little effect in the simulations of \cite{izidoro17}, mainly because of the modestly-higher planet masses. This put the super-Earths firmly in the inward migration regime, above the outward-migration zone depicted in Fig~\ref{fig:kep36_migmap}. However, \cite{raymond16} proposed that Jupiter's core formed close-in and migrated outward.} This quickly triggered a collision between the two innermost icy embryos to form a $7.3 \mearth$ icy planet. This increase in mass was sufficient to push the planet back toward inward migration. All of the other planets felt positive torques. The outer planets migrated out toward the zero-torque location, but the innermost rocky planet's outward migration was blocked by the inward-migrating large icy planet. This convergent migration pushed the planets toward successively-closer resonances, and they ended up locked in 7:6 resonance. This was demonstrated by the libration of the resonant angle $\Theta_1 = 7 \ \lambda_2 - 6 \ \lambda_1 - \varpi_1$, where $\lambda_1$ and $\lambda_2$ are the inner and outer planets' mean longitudes, respectively, and $\varpi_1$ is the inner planet's longitude of pericenter. The planets survived in resonance after the disk dissipated, to the end of the simulation. Meanwhile, the three outward-migrating icy embryos reached the zero-torque radius and ended up in a 1:2:3 resonant chain.  Two additional rocky embryos were shepherded outward and eventually collided with the icy embryos. 

The planets' feeding zones can be inferred from Fig.~\ref{fig:kep36_evol}, as all of the building blocks of each surviving planet have the same color.  The rocky planet formed from the collisional growth of nine terrestrial embryos, all of which originated between 0.7 and 1.1 AU.  Most of the four surviving icy planets' masses originated in the icy region from 5-10 AU. Three of the four accreted 1-2 rocky embryos but these accounted for just 3-6\% of their final masses.  

The two inner planets present an intriguing analog to the Kepler-36 system.  The Kepler-36 system contains two planets with well-constrained radii and masses~\citep{carter12}.  Kepler-36 b has a mass of $4.5 \mearth$ and a radius of $1.49 \rearth$ for a mean density of 7.46 $\gcm$, clearly within the rocky regime and consistent with the same composition as Earth.  In contrast, Kepler-36 c has a mass of $8 \mearth$ and a radius of $3.7 \rearth$ for a bulk density of 0.9 $\gcm$, indicating that it must have a significant gaseous envelope accounting for 1-10\% of its mass~\citep{carter12}.  The planets' orbital radii are 0.115 and 0.128 AU and they have a period ratio of 1.173, placing them very close to the 7:6 resonance. Numerical integrations suggest that their orbits are chaotic with a very short Lyapunov time~\citep{deck12}.  

Our two close-in simulated planets have drastically different feeding zones that may, by analogy, explain the very different radii of the Kepler-36 planets. The inner (lower-mass) planet is purely rocky and should have a composition and bulk density similar to Earth, as is the case for Kepler-36 b. The outer (higher-mass) simulated planet is mostly icy. While we do not model gas accretion and evaporation or radius evolution, the outer planet is more likely than the inner planet to have accreted a gaseous envelope because of its larger mass and longer time spent embedded within the disk, especially when the disk was at its densest~\citep[see, e.g., ][]{ikoma01}.

Our simulations also provide a decent match to the orbital architecture of the system. The simulated planets are locked in 7:6 resonance but their exact orbital period oscillates between 1.158 and 1.174, encompassing the Kepler-36 value of 1.173 and spending $\sim 20\%$ of their time within 0.001 of the true value. There are some small differences: the simulated planets are 30\% farther from their star, with the innermost planet at 0.15 AU. The mass ratio of our simulated planets is 2.4, slightly higher than the real system's value of 1.8. 

The origin of the density difference between the Kepler-36 planets has been attributed to photo-evaporation, which is much more efficient for planets with lower-mass cores~\citep{lopez13}. Yet we have shown that drastic density differences between adjacent planets can occur as a result of migration within the disk, with orbital configurations just as compact as the Kepler-36 system.  

\section{Diversity of outcomes}
\begin{figure}
  \leavevmode \epsfxsize=8cm\epsfbox{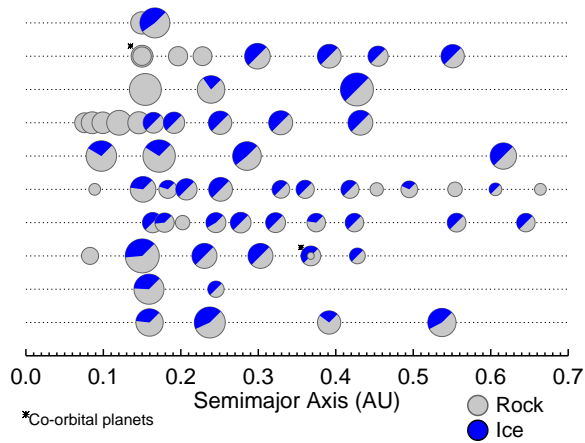}
    \caption[]{Final configuration of ten simulations illustrating the range of outcomes. Each planet's colors represent its rough composition: grey indicates rock and blue represents ice. Embryos that started past 5 AU started as 50-50 rock-ice mixtures and those from inside 5 AU were purely rocky. We do not account for various water loss processes and so the ice contents of simulated planets are certainly overestimates. The sizes of planets are scaled to their mass$^{1/3}$. The Kepler-36 analog system from Section 3 is at the top. Two co-orbital systems are marked with an asterisk. } 
     \label{fig:compare}
    \end{figure}

Figure~\ref{fig:compare} shows the final configuration of 10 simulations that illustrate the diversity of outcomes produced by the migration model. Here we have assumed that objects originating past the snow line (at 5 AU at the start of the simulations) are 50-50 rock-ice mixtures and those originating closer-in are pure rock. In many of these systems the innermost planets are purely rocky.  This emphasizes our main point that even if icy embryos migrate inward the innermost planets may still be rocky. In one system the five innermost planets are rocky. Of course, in many systems the innermost planet is dominated by mass by an embryo that migrated inward from past the snow line. In some systems rocky planets survived on orbits exterior to the icy embryos (Fig.~\ref{fig:compare}). In some long resonant chains, rocky planets are scattered throughout the chain. In systems with more widely-spaced orbits rocky planets are less common.  Most of these systems formed resonant chains that later became unstable~\citep[as in][]{izidoro17} and the rocky planets collided with icy ones. Two of the systems from Fig.~\ref{fig:compare} contain co-orbital planets, a common outcome of migration~\citep{cresswell09}. 

The water contents of simulated planets from Fig.~\ref{fig:compare} are likely overestimated. While cosmic abundance would suggest that icy embryos should form as 50-50 ice-rock mixtures~\citep{lodders03}, thermal and collisional processes may reduce the ice fraction. The constituent planetesimals of migrating ice-rich embryos form very early, leading to strong heating from short-lived radionuclides and dehydration~\citep{grimm93,monteux18}. Giant impacts may preferentially strip outer, icy mantles or surface oceans~\citep{genda05,marcus10}. However, upon inspection of the impacts suffered by super-Earths in our simulations we find that most impacts were relatively gentle, with collision speeds just above the two-body escape speed. Only a very small fraction of impacts was energetic enough to strip more than 10-20\% of the super-Earths' water~\citep[following ][]{marcus10}. This implies that late collisional losses are not likely to drastically alter the rock/water ratio.

\section{Limitations of our work} 
Our simulations are admittedly simple. There are large uncertainties \citep[and likely a large variety; ][]{bate18} in the structure and evolution of planet-forming disks~\citep[see][]{morbyraymond16}. We have chosen a single, albeit reasonable disk model~\citep{bitsch15}. We have not accounted for the fact that the disk's water content itself has a strong impact on embryos' migration and accretion histories~\citep{bitsch16}. Our simulations do not include processes that may be important for the formation of super-Earths, notably pebble accretion~\citep{ormel10,lambrechts12} and gas accretion/erosion~\citep{ginzburg16}. Our initial conditions are reasonable extrapolations of current models of embryo growth~\citep{kokubo02,morby15} but those models are themselves highly dependent on the previous phases of planetary formation, notably dust growth, drift, and planetesimal formation~\citep{johansen14,birnstiel16}.  It is a combination of all these processes that determine the relative abundance of rocky and icy embryos in the disk (including both when and where they form), and thus which embryos migrate and the final compositions of the super-Earths that form. Given our limited exploration of parameter space, the true diversity in super-Earth systems may be much larger than we have found.

\section{Discussion}

The prevalence of rocky vs icy super-Earths is a potential discriminant of planet formation models~\citep{raymond08a}. Of course, the closest-in planets are easiest to characterize~\citep[via bulk density measurements; e.g.,][]{marcy14}. Planets in the `photo-evaporation valley', at small masses very close to their stars, have recently been interpreted as being rocky in nature~\citep{owen17,lopez17,jin18}. The dip in planet occurrence between $\sim1.5$ and $2 \rearth$~\citep{fulton17} has also been interpreted as an indicator that many close-in planets are rocky~\citep{vaneylen17,jin18}. 

Our results serve to inject caution into the discussion. We have demonstrated that the inward migration of ice-rich embryos often acts to catalyze the growth of pure rock planets. The existence of very close-in rocky super-Earths cannot be used as evidence against the migration model. It is even possible that planets originating past the snow line contain far less water than generally assumed. Strong radiogenic heating should completely dehydrate fast-growing planetesimals~\citep{grimm93,monteux18}, yet planetesimals past the snow line should continue to accrete icy pebbles. The giant impacts in our simulations were not energetic enough to produce strong water/ice loss. We conclude that further studies are needed to assess the water retention/loss on super-Earths that form beyond the snow line. Nonetheless, it is hard to imagine the migration model producing {\it only} rocky super-Earths, unless a mechanism exists to preferentially produce large (migrating) rocky embryos rather than icy ones.

Our Kepler-36 analog simulation illustrates how adjacent planets can have drastically different feeding zones. In the face of migration, planets' feeding zones are often very wide and disconnected from their final orbital locations. The rocky Kepler-36 b analog accreted entirely from embryos between 0.7 and 1.1 AU but ended up at 0.15 AU (see Fig.~\ref{fig:kep36_evol}). Likewise, the ice-rich Kepler-36 c analog finished the simulation at 0.17 AU but accreted mainly from embryos that originated past 5 AU. The large density difference between the Kepler-36 planets may plausibly be explained a result of the system's formation process.

\vskip .2in
\noindent {\em Acknowledgments.}  We thank the referee for a constructive report. B.~B. thanks the European Research Council (ERC Starting Grant 757448-PAMDORA) for their financial support. A.~I. and L.~E. are grateful to FAPESP for financial support through grants 16/12686-2 and 16/19556-7 (A.~I.) and 2017/09963-7 (L.~E.). S.~N.~R. thanks the Agence Nationale pour la Recherche via grant ANR-13-BS05-0003-002 ({\it MOJO}).


\begin{thebibliography}{}

\bibitem[\protect\citeauthoryear{{Baraffe}, {Alibert}, {Chabrier} \&
  {Benz}}{{Baraffe} et~al.}{2006}]{baraffe06}
{Baraffe} I.,  {Alibert} Y.,  {Chabrier} G.,    {Benz} W.,  2006, \aap, 450,
  1221

\bibitem[\protect\citeauthoryear{{Batalha} et~al.,}{{Batalha}
  et~al.}{2013}]{batalha13}
{Batalha} N.~M.  et~al., 2013, \apjs, 204, 24

\bibitem[\protect\citeauthoryear{{Bate}}{{Bate}}{2018}]{bate18}
{Bate} M.~R.,  2018, \mnras, 475, 5618

\bibitem[\protect\citeauthoryear{{Birnstiel}, {Fang} \& {Johansen}}{{Birnstiel}
  et~al.}{2016}]{birnstiel16}
{Birnstiel} T.,  {Fang} M.,    {Johansen} A.,  2016, \ssr, 205, 41

\bibitem[\protect\citeauthoryear{{Bitsch}, {Crida}, {Morbidelli}, {Kley} \&
  {Dobbs-Dixon}}{{Bitsch} et~al.}{2013}]{bitsch13}
{Bitsch} B.,  {Crida} A.,  {Morbidelli} A.,  {Kley} W.,    {Dobbs-Dixon} I.,
  2013, \aap, 549, A124

\bibitem[\protect\citeauthoryear{{Bitsch} \& {Johansen}}{{Bitsch} \&
  {Johansen}}{2016}]{bitsch16}
{Bitsch} B.,  {Johansen} A.,  2016, \aap, 590, A101

\bibitem[\protect\citeauthoryear{{Bitsch}, {Johansen}, {Lambrechts} \&
  {Morbidelli}}{{Bitsch} et~al.}{2015}]{bitsch15}
{Bitsch} B.,  {Johansen} A.,  {Lambrechts} M.,    {Morbidelli} A.,  2015, \aap,
  575, A28

\bibitem[\protect\citeauthoryear{{Bitsch} \& {Kley}}{{Bitsch} \&
  {Kley}}{2010}]{bitsch10}
{Bitsch} B.,  {Kley} W.,  2010, \aap, 523, A30

\bibitem[\protect\citeauthoryear{{Bitsch}, {Morbidelli}, {Lega} \&
  {Crida}}{{Bitsch} et~al.}{2014}]{bitsch14}
{Bitsch} B.,  {Morbidelli} A.,  {Lega} E.,    {Crida} A.,  2014, \aap, 564,
  A135

\bibitem[\protect\citeauthoryear{{Boley} \& {Ford}}{{Boley} \&
  {Ford}}{2013}]{boley13}
{Boley} A.~C.,  {Ford} E.~B.,  2013, arXiv:1306.0566

\bibitem[\protect\citeauthoryear{{Bolmont}, {Raymond}, {von Paris}, {Selsis},
  {Hersant}, {Quintana} \& {Barclay}}{{Bolmont} et~al.}{2014}]{bolmont14}
{Bolmont} E.,  {Raymond} S.~N.,  {von Paris} P.,  {Selsis} F.,  {Hersant} F.,
  {Quintana} E.~V.,    {Barclay} T.,  2014, \apj, 793, 3

\bibitem[\protect\citeauthoryear{{Carrera}, {Ford}, {Izidoro}, {Jontof-Hutter},
  {Raymond} \& {Wolfgang}}{{Carrera} et~al.}{2018}]{carrera18}
{Carrera} D.,  {Ford} E.~B.,  {Izidoro} A.,  {Jontof-Hutter} D.,  {Raymond}
  S.~N.,    {Wolfgang} A.,  2018, ArXiv e-prints

\bibitem[\protect\citeauthoryear{{Carter} et~al.,}{{Carter}
  et~al.}{2012}]{carter12}
{Carter} J.~A.  et~al., 2012, Science, 337, 556

\bibitem[\protect\citeauthoryear{{Chambers}}{{Chambers}}{1999}]{chambers99}
{Chambers} J.~E.,  1999, \mnras, 304, 793

\bibitem[\protect\citeauthoryear{{Chatterjee} \& {Tan}}{{Chatterjee} \&
  {Tan}}{2014}]{chatterjee14}
{Chatterjee} S.,  {Tan} J.~C.,  2014, \apj, 780, 53

\bibitem[\protect\citeauthoryear{{Chatterjee} \& {Tan}}{{Chatterjee} \&
  {Tan}}{2015}]{chatterjee15}
{Chatterjee} S.,  {Tan} J.~C.,  2015, \apjl, 798, L32

\bibitem[\protect\citeauthoryear{{Cossou}, {Raymond}, {Hersant} \&
  {Pierens}}{{Cossou} et~al.}{2014}]{cossou14}
{Cossou} C.,  {Raymond} S.~N.,  {Hersant} F.,    {Pierens} A.,  2014, \aap,
  569, A56

\bibitem[\protect\citeauthoryear{{Cossou}, {Raymond} \& {Pierens}}{{Cossou}
  et~al.}{2013}]{cossou13}
{Cossou} C.,  {Raymond} S.~N.,    {Pierens} A.,  2013, \aap, 553, L2

\bibitem[\protect\citeauthoryear{{Cresswell} \& {Nelson}}{{Cresswell} \&
  {Nelson}}{2008}]{cresswell08}
{Cresswell} P.,  {Nelson} R.~P.,  2008, \aap, 482, 677

\bibitem[\protect\citeauthoryear{{Cresswell} \& {Nelson}}{{Cresswell} \&
  {Nelson}}{2009}]{cresswell09}
{Cresswell} P.,  {Nelson} R.~P.,  2009, \aap, 493, 1141

\bibitem[\protect\citeauthoryear{{Deck}, {Holman}, {Agol}, {Carter},
  {Lissauer}, {Ragozzine} \& {Winn}}{{Deck} et~al.}{2012}]{deck12}
{Deck} K.~M.,  {Holman} M.~J.,  {Agol} E.,  {Carter} J.~A.,  {Lissauer} J.~J.,
  {Ragozzine} D.,    {Winn} J.~N.,  2012, \apjl, 755, L21

\bibitem[\protect\citeauthoryear{{Fabrycky} et~al.,}{{Fabrycky}
  et~al.}{2014}]{fabrycky14}
{Fabrycky} D.~C.  et~al., 2014, \apj, 790, 146

\bibitem[\protect\citeauthoryear{{Fendyke} \& {Nelson}}{{Fendyke} \&
  {Nelson}}{2014}]{fendyke14}
{Fendyke} S.~M.,  {Nelson} R.~P.,  2014, \mnras, 437, 96

\bibitem[\protect\citeauthoryear{{Fogg} \& {Nelson}}{{Fogg} \&
  {Nelson}}{2005}]{fogg05}
{Fogg} M.~J.,  {Nelson} R.~P.,  2005, \aap, 441, 791

\bibitem[\protect\citeauthoryear{{Fulton} et~al.,}{{Fulton}
  et~al.}{2017}]{fulton17}
{Fulton} B.~J.  et~al., 2017, \aj, 154, 109

\bibitem[\protect\citeauthoryear{{Genda} \& {Abe}}{{Genda} \&
  {Abe}}{2005}]{genda05}
{Genda} H.,  {Abe} Y.,  2005, \nat, 433, 842

\bibitem[\protect\citeauthoryear{{Ginzburg}, {Schlichting} \&
  {Sari}}{{Ginzburg} et~al.}{2016}]{ginzburg16}
{Ginzburg} S.,  {Schlichting} H.~E.,    {Sari} R.,  2016, \apj, 825, 29

\bibitem[\protect\citeauthoryear{{Grimm} \& {McSween}}{{Grimm} \&
  {McSween}}{1993}]{grimm93}
{Grimm} R.~E.,  {McSween} H.~Y.,  1993, Science, 259, 653

\bibitem[\protect\citeauthoryear{{Hubbard}, {Hattori}, {Burrows}, {Hubeny} \&
  {Sudarsky}}{{Hubbard} et~al.}{2007}]{hubbard07b}
{Hubbard} W.~B.,  {Hattori} M.~F.,  {Burrows} A.,  {Hubeny} I.,    {Sudarsky}
  D.,  2007, Icarus, 187, 358

\bibitem[\protect\citeauthoryear{{Ida} \& {Lin}}{{Ida} \& {Lin}}{2010}]{ida10}
{Ida} S.,  {Lin} D.~N.~C.,  2010, \apj, 719, 810

\bibitem[\protect\citeauthoryear{{Ikoma}, {Emori} \& {Nakazawa}}{{Ikoma}
  et~al.}{2001}]{ikoma01}
{Ikoma} M.,  {Emori} H.,    {Nakazawa} K.,  2001, \apj, 553, 999

\bibitem[\protect\citeauthoryear{{Inamdar} \& {Schlichting}}{{Inamdar} \&
  {Schlichting}}{2015}]{inamdar15}
{Inamdar} N.~K.,  {Schlichting} H.~E.,  2015, \mnras, 448, 1751

\bibitem[\protect\citeauthoryear{{Izidoro}, {Morbidelli} \&
  {Raymond}}{{Izidoro} et~al.}{2014}]{izidoro14}
{Izidoro} A.,  {Morbidelli} A.,    {Raymond} S.~N.,  2014, \apj, 794, 11

\bibitem[\protect\citeauthoryear{{Izidoro}, {Ogihara}, {Raymond}, {Morbidelli},
  {Pierens}, {Bitsch}, {Cossou} \& {Hersant}}{{Izidoro}
  et~al.}{2017}]{izidoro17}
{Izidoro} A.,  {Ogihara} M.,  {Raymond} S.~N.,  {Morbidelli} A.,  {Pierens} A.,
   {Bitsch} B.,  {Cossou} C.,    {Hersant} F.,  2017, \mnras, 470, 1750

\bibitem[\protect\citeauthoryear{{Jin} \& {Mordasini}}{{Jin} \&
  {Mordasini}}{2018}]{jin18}
{Jin} S.,  {Mordasini} C.,  2018, \apj, 853, 163

\bibitem[\protect\citeauthoryear{{Johansen}, {Blum}, {Tanaka}, {Ormel},
  {Bizzarro} \& {Rickman}}{{Johansen} et~al.}{2014}]{johansen14}
{Johansen} A.,  {Blum} J.,  {Tanaka} H.,  {Ormel} C.,  {Bizzarro} M.,
  {Rickman} H.,  2014, Protostars and Planets VI, pp 547--570

\bibitem[\protect\citeauthoryear{{Kokubo} \& {Ida}}{{Kokubo} \&
  {Ida}}{2002}]{kokubo02}
{Kokubo} E.,  {Ida} S.,  2002, \apj, 581, 666

\bibitem[\protect\citeauthoryear{{Lambrechts} \& {Johansen}}{{Lambrechts} \&
  {Johansen}}{2012}]{lambrechts12}
{Lambrechts} M.,  {Johansen} A.,  2012, \aap, 544, A32

\bibitem[\protect\citeauthoryear{{Lammer}, {Selsis}, {Ribas}, {Guinan}, {Bauer}
  \& {Weiss}}{{Lammer} et~al.}{2003}]{lammer03}
{Lammer} H.,  {Selsis} F.,  {Ribas} I.,  {Guinan} E.~F.,  {Bauer} S.~J.,
  {Weiss} W.~W.,  2003, \apjl, 598, L121

\bibitem[\protect\citeauthoryear{{Lissauer} et~al.,}{{Lissauer}
  et~al.}{2011}]{lissauer11b}
{Lissauer} J.~J.  et~al., 2011, \apjs, 197, 8

\bibitem[\protect\citeauthoryear{{Lodders}}{{Lodders}}{2003}]{lodders03}
{Lodders} K.,  2003, \apj, 591, 1220

\bibitem[\protect\citeauthoryear{{Lopez}}{{Lopez}}{2017}]{lopez17}
{Lopez} E.~D.,  2017, \mnras, 472, 245

\bibitem[\protect\citeauthoryear{{Lopez} \& {Fortney}}{{Lopez} \&
  {Fortney}}{2013}]{lopez13}
{Lopez} E.~D.,  {Fortney} J.~J.,  2013, \apj, 776, 2

\bibitem[\protect\citeauthoryear{{Mandell}, {Raymond} \&
  {Sigurdsson}}{{Mandell} et~al.}{2007}]{mandell07}
{Mandell} A.~M.,  {Raymond} S.~N.,    {Sigurdsson} S.,  2007, \apj, 660, 823

\bibitem[\protect\citeauthoryear{{Marcus}, {Sasselov}, {Stewart} \&
  {Hernquist}}{{Marcus} et~al.}{2010}]{marcus10}
{Marcus} R.~A.,  {Sasselov} D.,  {Stewart} S.~T.,    {Hernquist} L.,  2010,
  \apjl, 719, L45

\bibitem[\protect\citeauthoryear{{Marcy} et~al.,}{{Marcy}
  et~al.}{2014}]{marcy14}
{Marcy} G.~W.  et~al., 2014, \apjs, 210, 20

\bibitem[\protect\citeauthoryear{{Masset}, {Morbidelli}, {Crida} \&
  {Ferreira}}{{Masset} et~al.}{2006}]{masset06}
{Masset} F.~S.,  {Morbidelli} A.,  {Crida} A.,    {Ferreira} J.,  2006, \apj,
  642, 478

\bibitem[\protect\citeauthoryear{{McNeil} \& {Nelson}}{{McNeil} \&
  {Nelson}}{2010}]{mcneil10}
{McNeil} D.~S.,  {Nelson} R.~P.,  2010, \mnras, 401, 1691

\bibitem[\protect\citeauthoryear{{Monteux}, {Golabek}, {Rubie}, {Tobie} \&
  {Young}}{{Monteux} et~al.}{2018}]{monteux18}
{Monteux} J.,  {Golabek} G.~J.,  {Rubie} D.~C.,  {Tobie} G.,    {Young} E.~D.,
  2018, \ssr, 214, 39

\bibitem[\protect\citeauthoryear{{Morbidelli}, {Lambrechts}, {Jacobson} \&
  {Bitsch}}{{Morbidelli} et~al.}{2015}]{morby15}
{Morbidelli} A.,  {Lambrechts} M.,  {Jacobson} S.,    {Bitsch} B.,  2015,
  Icarus, 258, 418

\bibitem[\protect\citeauthoryear{{Morbidelli} \& {Raymond}}{{Morbidelli} \&
  {Raymond}}{2016}]{morbyraymond16}
{Morbidelli} A.,  {Raymond} S.~N.,  2016, Journal of Geophysical Research
  (Planets), 121, 1962

\bibitem[\protect\citeauthoryear{{Ogihara} \& {Ida}}{{Ogihara} \&
  {Ida}}{2009}]{ogihara09}
{Ogihara} M.,  {Ida} S.,  2009, \apj, 699, 824

\bibitem[\protect\citeauthoryear{{Ogihara}, {Morbidelli} \&
  {Guillot}}{{Ogihara} et~al.}{2015}]{ogihara15}
{Ogihara} M.,  {Morbidelli} A.,    {Guillot} T.,  2015, \aap, 578, A36

\bibitem[\protect\citeauthoryear{{Ormel} \& {Klahr}}{{Ormel} \&
  {Klahr}}{2010}]{ormel10}
{Ormel} C.~W.,  {Klahr} H.~H.,  2010, \aap, 520, A43

\bibitem[\protect\citeauthoryear{{Owen} \& {Wu}}{{Owen} \& {Wu}}{2013}]{owen13}
{Owen} J.~E.,  {Wu} Y.,  2013, \apj, 775, 105

\bibitem[\protect\citeauthoryear{{Owen} \& {Wu}}{{Owen} \& {Wu}}{2017}]{owen17}
{Owen} J.~E.,  {Wu} Y.,  2017, \apj, 847, 29

\bibitem[\protect\citeauthoryear{{Paardekooper}, {Baruteau}, {Crida} \&
  {Kley}}{{Paardekooper} et~al.}{2010}]{paardekooper10}
{Paardekooper} S.-J.,  {Baruteau} C.,  {Crida} A.,    {Kley} W.,  2010, \mnras,
  401, 1950

\bibitem[\protect\citeauthoryear{{Paardekooper}, {Baruteau} \&
  {Kley}}{{Paardekooper} et~al.}{2011}]{paardekooper11}
{Paardekooper} S.-J.,  {Baruteau} C.,    {Kley} W.,  2011, \mnras, 410, 293

\bibitem[\protect\citeauthoryear{{Papaloizou} \& {Larwood}}{{Papaloizou} \&
  {Larwood}}{2000}]{papaloizou00}
{Papaloizou} J.~C.~B.,  {Larwood} J.~D.,  2000, \mnras, 315, 823

\bibitem[\protect\citeauthoryear{{Raymond}, {Barnes} \& {Mandell}}{{Raymond}
  et~al.}{2008}]{raymond08a}
{Raymond} S.~N.,  {Barnes} R.,    {Mandell} A.~M.,  2008, \mnras, 384, 663

\bibitem[\protect\citeauthoryear{{Raymond}, {Izidoro}, {Bitsch} \&
  {Jacobson}}{{Raymond} et~al.}{2016}]{raymond16}
{Raymond} S.~N.,  {Izidoro} A.,  {Bitsch} B.,    {Jacobson} S.~A.,  2016,
  \mnras, 458, 2962

\bibitem[\protect\citeauthoryear{{Raymond}, {Kokubo}, {Morbidelli}, {Morishima}
  \& {Walsh}}{{Raymond} et~al.}{2014}]{raymond14}
{Raymond} S.~N.,  {Kokubo} E.,  {Morbidelli} A.,  {Morishima} R.,    {Walsh}
  K.~J.,  2014, Protostars and Planets VI, pp 595--618

\bibitem[\protect\citeauthoryear{{Raymond}, {Mandell} \&
  {Sigurdsson}}{{Raymond} et~al.}{2006}]{raymond06c}
{Raymond} S.~N.,  {Mandell} A.~M.,    {Sigurdsson} S.,  2006, Science, 313,
  1413

\bibitem[\protect\citeauthoryear{{Romanova} \& {Lovelace}}{{Romanova} \&
  {Lovelace}}{2006}]{romanova06}
{Romanova} M.~M.,  {Lovelace} R.~V.~E.,  2006, \apjl, 645, L73

\bibitem[\protect\citeauthoryear{{Sanchis-Ojeda}, {Rappaport}, {Winn},
  {Kotson}, {Levine} \& {El Mellah}}{{Sanchis-Ojeda}
  et~al.}{2014}]{sanchisojeda14}
{Sanchis-Ojeda} R.,  {Rappaport} S.,  {Winn} J.~N.,  {Kotson} M.~C.,  {Levine}
  A.,    {El Mellah} I.,  2014, \apj, 787, 47

\bibitem[\protect\citeauthoryear{{Tanaka} \& {Ward}}{{Tanaka} \&
  {Ward}}{2004}]{tanaka04}
{Tanaka} H.,  {Ward} W.~R.,  2004, \apj, 602, 388

\bibitem[\protect\citeauthoryear{{Terquem} \& {Papaloizou}}{{Terquem} \&
  {Papaloizou}}{2007}]{terquem07}
{Terquem} C.,  {Papaloizou} J.~C.~B.,  2007, \apj, 654, 1110

\bibitem[\protect\citeauthoryear{{Van Eylen}, {Agentoft}, {Lundkvist},
  {Kjeldsen}, {Owen}, {Fulton}, {Petigura} \& {Snellen}}{{Van Eylen}
  et~al.}{2017}]{vaneylen17}
{Van Eylen} V.,  {Agentoft} C.,  {Lundkvist} M.~S.,  {Kjeldsen} H.,  {Owen}
  J.~E.,  {Fulton} B.~J.,  {Petigura} E.,    {Snellen} I.,  2017,
  arxiv:1710.05398

\end{thebibliography}

\end{document}